# First demonstration of THGEM/GAPD-matrix optical readout in two-phase Cryogenic Avalanche Detector in Ar

A. Bondar [a,b], A. Buzulutskov [a,b]*, A. Dolgov [b], A. Grebenuk [a], E. Shemyakina [a,b], A. Sokolov [a,b], A. Breskin [c], D. Thers [d]

[a] *Budker Institute of Nuclear Physics, 630090 Novosibirsk, Russia*
[b] *Novosibirsk State University, Pirogov street 2, 630090 Novosibirsk, Russia*
[c] *Weizmann Institute of Science, 76100 Rehovot, Israel*
[d] *SUBATECH, Ecole des Mines de Nantes, CNRS/In2p3, Universite de Nantes, 44307 Nantes Cedex 3, France*

**Abstract**

The multi-channel optical readout of a THGEM multiplier coupled to a matrix of 3x3 Geiger-mode APDs (GAPDs) was demonstrated in a two-phase Cryogenic Avalanche Detector (CRAD) in Ar. The GAPDs recorded THGEM-hole avalanches in the Near Infrared (NIR). At an avalanche charge gain of 160, the yield of the combined THGEM/GAPD-matrix multiplier amounted at ~80 photoelectrons per 20 keV X-ray absorbed in the liquid phase. A spatial resolution of 2.5 mm (FWHM) has been measured for the impinging X-rays. This technique has potential applications in coherent neutrino-nucleus scattering and dark matter search experiments.

## 1. Introduction

In two-phase Cryogenic Avalanche Detectors (CRADs), the primary-ionization electrons produced in the noble liquid are emitted into the gas phase and can be multiplied with Gas Electron Multipliers (GEMs), thick GEMs (THGEMs) or other elements (see review [1]). The multiplied electrons can be recorded through direct charge readout mode, or through optical recording of avalanche photons, e.g. with THGEMs [2] coupled to Geiger-mode APDs (GAPDs) [3]. The final goal for this kind of devices is the development of large-volume detectors of ultimate sensitivity for rare-event experiments; examples are coherent neutrino-nucleus scattering, direct dark matter searches and astrophysical neutrino detection experiments [1].

In terms of the maximum charge gain, most promising results have been obtained in two-phase CRADs operated in Ar, with gains reaching 1000 and 5000 with the double-THGEM [4],[5] and hybrid double-THGEM/GEM [5] multipliers respectively. These results however are not sufficient for effective operation with ultimate sensitivity, i.e. in single-electron counting mode and with precise 2D readout.

A potential solution might be a combined THGEM/GAPD multiplier [6],[7],[8], with optical recording of THGEM-hole avalanches either in the VUV [6],[8] or in Near Infrared (NIR) [7],[9]. Here the avalanche gain can be substantially reduced, compensated by the high GAPD gain. The 2D readout can be provided by a matrix of GAPDs placed behind the THGEM multiplier [1],[8],[9]. In previous works such an optical readout was studied either in Ar, with single-channel GAPDs [6],[7], or in Xe, albeit with the multi-channel GAPD array but without presenting the experimental coordinate characteristics of the detector [8].

In this work, the multi-channel optical readout in a two-phase CRAD in Ar is demonstrated for the first time with a THGEM/GAPD-matrix multiplier in terms of a high spatial resolution; more elaborated results will be presented elsewhere [10].

## 2. Experimental setup

The experimental setup (see Figs. 1 and 2) was similar to that used in our previous measurements [5],[7]. It includes a cryostat with a 9 l volume cryogenic chamber. The chamber comprised of a cathode mesh, immersed in a ~1 cm thick liquid-Ar layer, and a double-THGEM assembly of an active area of $10\times10$ cm$^2$, placed in the gas phase above the liquid. The detector was operated in two-phase mode in equilibrium state, at a saturated vapour pressure of 1.0 atm and temperature of 87 K. The Ar was purified by an Oxisorb filter, providing an electron life-time of >13 μs in the liquid. The THGEM geometrical parameters were the same as in [5]. The cathode and THGEM electrodes were biased through a resistive high-voltage divider, placed outside the cryostat. In all the measurements, the electric field within liquid Ar was kept at 1.76 kV/cm.

* Corresponding author. *E-mail address*: a.f.buzulutskov@inp.nsk.su.

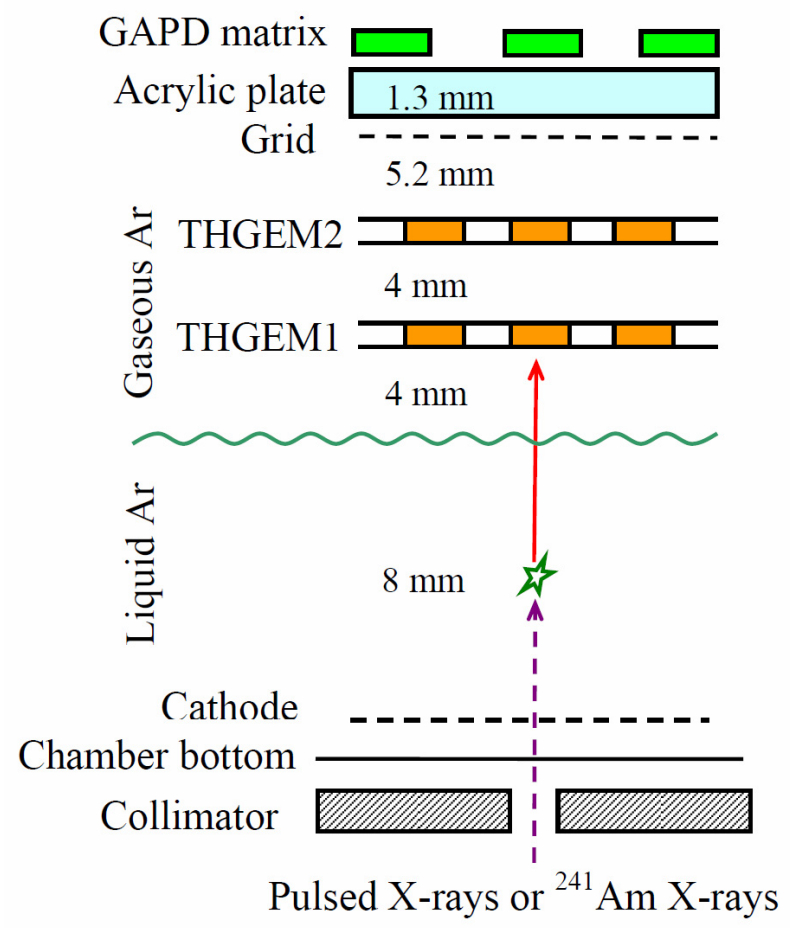

Fig. 1. Experimental setup.

A 3x3 matrix of GAPDs (MRS APD "CPTA 149-35") was placed in the gas phase at a distance of 6.5 mm behind the second THGEM, electrically insulated from the latter with an acrylic plate transparent in the NIR and by a wire grid at ground potential (Fig. 2). Each GAPD had a 2.1×2.1 $mm^2$ active area and a Photon Detection Efficiency (PDE) of about 15% at 800 nm [7].

The detector was irradiated from outside through aluminium windows in two ways: (a) by 15-40 keV X-rays from a pulsed X-ray tube (at a rate of 240 Hz) through a cylindrical collimator with a hole diameter of 2 mm; (b) by X-rays from a $^{241}$Am source (a 60 keV and lower energy lines, at a rate of a few Hz), through another cylindrical collimator with a hole diameter of 15 mm.

In case of $^{241}$Am X-rays, the source was placed at a distance of 2.4 cm from the collimator output, providing a broad irradiation area in liquid Ar, of about 15 mm in diameter.

In case of pulsed X-rays, the X-ray tube was placed at a relatively large distance, of ~50 cm, from the collimator; it permitted an operation in counting mode, with a rather small deposited energy per pulse, of 20 keV on average. Here, the X-ray conversion in the liquid occured over a small area, of ~2 mm in diameter, thus allowing for accurate estimate of the spatial resolution.

The charge signals were recorded from the last electrode of the second THGEM using a charge-sensitive amplifier. The optical signals were recorded from the GAPDs via twisted-pair cables connected to fast amplifiers with 300 MHz bandwidth and an amplification factor of 30. The DAQ system included an 8-channel Flash ADC CAEN V1720 (12 bit, 250 MHz): the signals from 7 GAPDs and from the THGEM were digitized and stored in a computer for further off-line analysis using LabView (resulting in that only 7 out of 9 GAPDs were active).

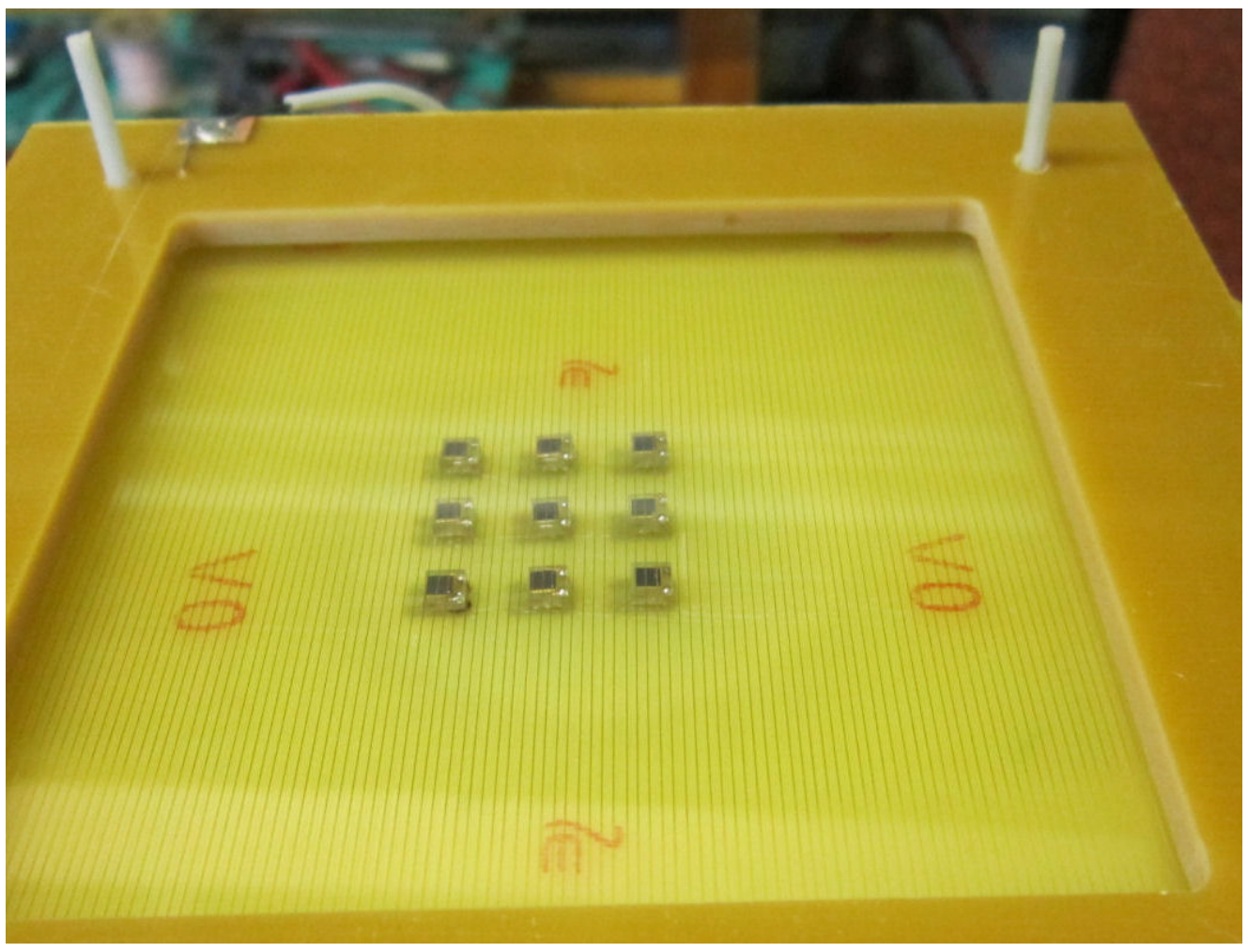

Fig. 2. Photograph of the GAPD-matrix, acrylic plate and wire grid assembly at the mounting stage of the two-phase CRAD.

In case of $^{241}$Am X-rays, a self-triggering mode was applied, provided by the double-THGEM signals exceeding a certain threshold. Some noise signals, not related to X-ray conversion in the liquid, were superimposed in this case. In case of pulsed X-rays, the trigger was external, provided by a pulsed X-tube generator. In this case the contribution of noise signals was negligible.

Other details of the experimental setup and measurement procedures can be found elsewhere [5],[7].

## 3. Results

Since the avalanche scintillations in the two-phase CRAD were observed from the THGEM holes using GAPDs without a wavelength shifter (WLS), i.e. insensitive to UV, only the Ar-emitted NIR photons were recorded, as established elsewhere [7],[9]. Fig. 3 illustrates a typical avalanche scintillation event, along with the performance of the multi-channel DAQ system: a typical event in the two-phase Ar CRAD is shown induced by a $^{241}$Am X-ray, at a double-THGEM gain of 260. The slow signals exhibited by those of the charge and optical channels, exceeding 20 μs, are due to the slow electron emission component in two-phase Ar systems [1]. The optical (GAPD) signals are composed of a number of fast pulses, each apparently corresponding to a single-pixel discharge (or those of multi-pixels in case of cross-talks), i.e. to the detection of a single photoelectron (pe). Accordingly, the amplitude of each GAPD signal, expressed in number of photoelectrons, was measured by counting the number of peaks presented in that signal using a dedicated peak-finder algorithm.

For infrequent signals induced by the $^{241}$Am source and those of GAPD noises, the GAPD-pulse amplitude distribution had a nominal shape: a well-defined single-pixel peak accompanied by secondary (cross-talk) peaks (see Fig. 4).

However at higher rates, when irradiated with the pulsed X-ray tube (240Hz), a specific GAPD saturation effect has been revealed: the pixel-pulse amplitude substantially decreased, degrading the amplitude spectrum. In this case, some pulses were even lost below threshold, thus reducing the PDE of the GAPD. This GAPD rate-dependence problem was studied elsewhere [11]: it is shown to be due to a considerable increase of the pixel quenching resistor at low temperatures, observed earlier for this particular GAPD type [7].

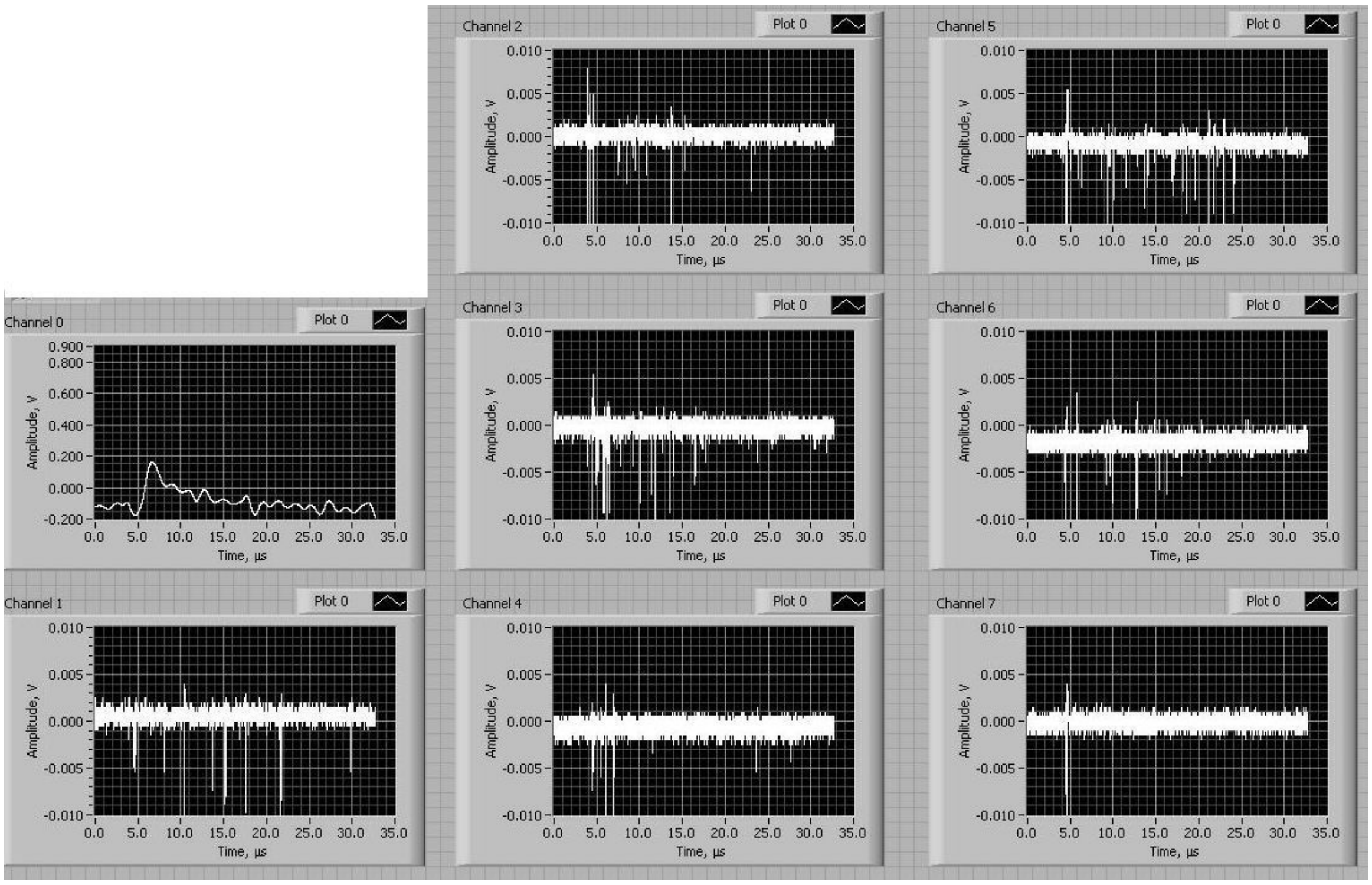

Fig. 3. A typical event recorded with the FADC in the two-phase Ar CRAD with the THGEM/GAPD-matrix optical readout (of Fig. 1), induced by a $^{241}$Am X-ray. Shown are the charge signal from the double-THGEM (channel 0) and the optical signals recorded from the GAPD matrix (channels 1-7). Double-THGEM gain: 260; THGEM amplifier shaping time: 0.5 µs; GAPDs bias voltage: 36 V. The trigger is provided by the double-THGEM signal.

Regarding the yield of the combined THGEM/GAPD-matrix multiplier in the two-phase Ar CRAD, it was measured accurately when irradiated with the pulsed X-rays. The total GAPD-matrix amplitude, for 7 active channels of the matrix, amounted at 80 pe per 20 keV X-ray, at a double-THGEM charge gain of 160 (here the amplitude of the central GAPD amounted to about 20 pe). That means that we may still have reasonable THGEM/GAPD-matrix yield for lower energy deposition: in particular, more than 10 pe are expected per 1 keV of deposited energy at a charge gain of 600, i.e. ≥10pe/keVee. Even higher yields might be expected when the GAPD rate problem will be solved.

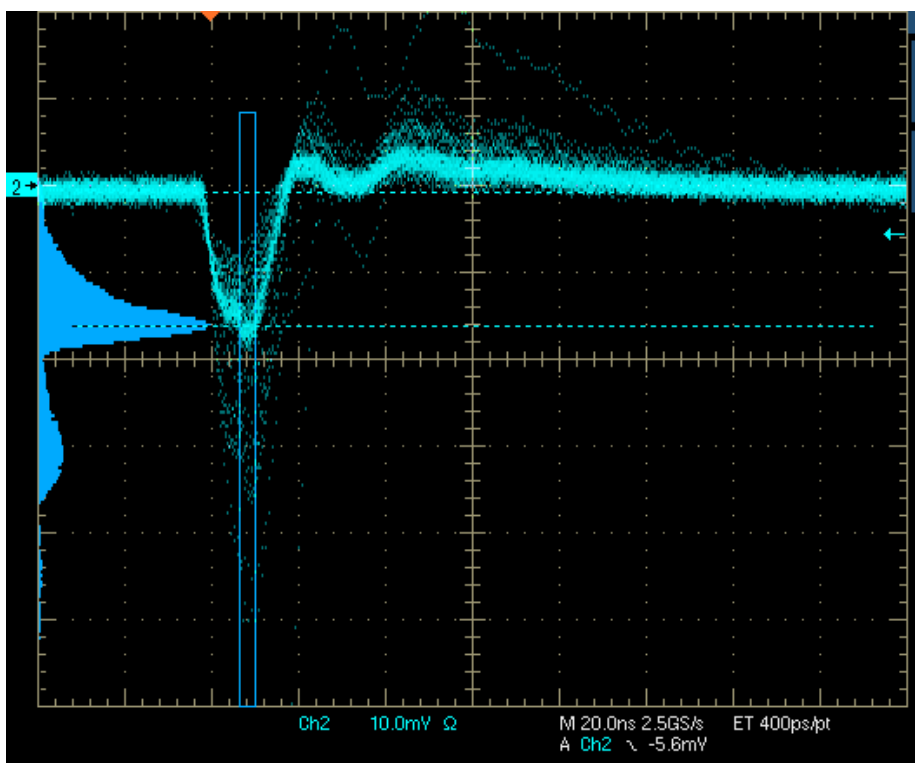

Fig. 4. GAPD noise pulses and their pulse-height distribution (on the left) of the GAPD #8 in the two-phase Ar CRAD with the THGEM/GAPD-matrix optical readout. Scales: 20 ns/div; 10 mV/div. GAPD bias voltage: 40 V.

The coordinate characteristics of the two-phase CRAD with the THGEM/GAPD-matrix readout are demonstrated in Figs. 5 and 6. Fig. 5 illustrates the imaging capability of the detector: the X-Y coordinate plots of the X-ray conversion regions are shown. The latter are defined by either 15 mm or 2 mm diameter collimator, for $^{241}$Am X-rays (and noise signals) and pulsed X-rays respectively.

The X-Y coordinates were calculated from the GAPDs amplitudes using a center-of-gravity algorithm, corrected for simulation of the light propagation from THGEM holes to a given GAPD. To determine the X or Y coordinate, the GAPD amplitudes from either the row or the column of the GAPD matrix were used. In the latter case, two Y-columns were used, which significantly improved the spatial resolution compared to that of a single column.

In case of $^{241}$Am X-rays, the circle in the figure outlines the 15 mm diameter collimator. The non-uniformity of the event population within the circle area is caused by a $^{241}$Am source position eccentricity with respect to the collimator. Also note that the events concentration in the vicinity of GAPD #5 is presumably induced by noise signals.

In general, one may conclude that the distribution shapes in Fig. 5 match the different collimators shapes, thus confirming the imaging capability of the detector. This is clearer in Fig. 6, showing the Y projective coordinate distribution in both cases. The width of each distribution, of 16 mm and 3.2 mm (FWHM), corresponds to the appropriate collimator diameters, of 15 mm and 2 mm. The spatial resolution of the detector can be estimated from the data of the smaller diameter beam, by quadratic subtraction; it results in a resolution of ~2.5 mm (FWHM) for 20 keV energy deposition, at the present operation conditions.

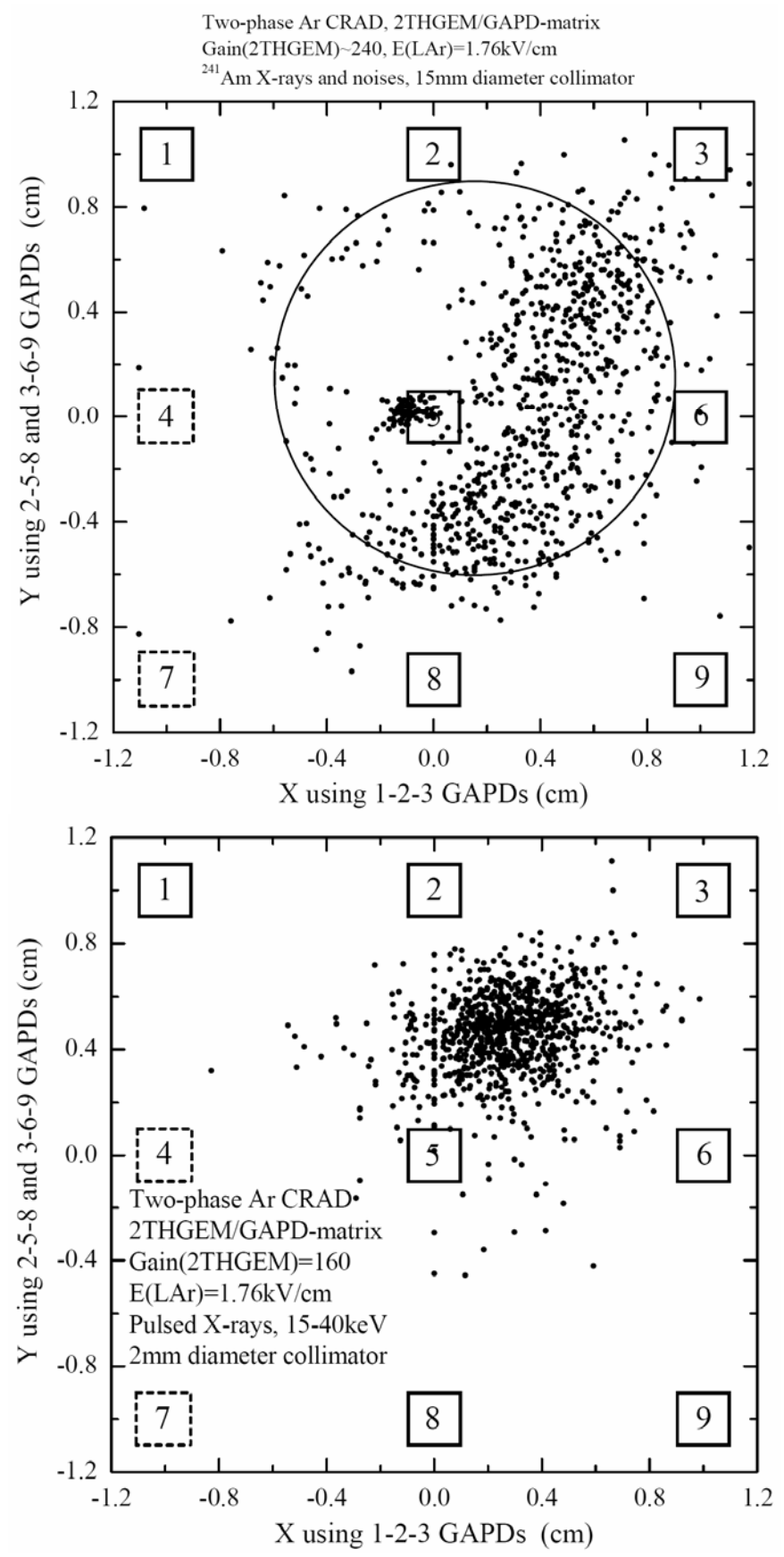

Fig. 5. X-Y plots of the X-ray conversion region in the two-phase Ar CRAD with THGEM/GAPD-matrix optical readout (of Fig. 1), defined by either 15 mm (top figure) or 2 mm (bottom figure) diameter collimator, for $^{241}$Am X-rays (+ noises) and pulsed X-rays, respectively. Squares with numbers indicate the positioning of the active areas of the GAPDs. The circle in the top figure outlines the 15 mm diameter collimator.

## 4. Conclusions

A multi-channel optical readout of a Cryogenic Avalanche Detector (CRAD) operating in Ar in a two-phase mode, was demonstrated for the first time - using a combined THGEM/GAPD-matrix multiplier. The matrix was composed of a 3x3 array of GAPDs, optically recording THGEM-hole avalanches in the NIR. At an avalanche charge gain of 160, the yield of the combined THGEM/GAPD-matrix multiplier amounted at ~80 photoelectrons per 20 keV X-ray absorbed in liquid Ar. The detector had a spatial resolution of 2.5 mm (FWHM), rather high for two-phase detectors (as compared for example to that of [12]), and a low detection threshold. More elaborated results will be presented elsewhere [10].

This type of detector may find applications in low-threshold rare-event experiments, such as coherent neutrino-nucleus scattering and dark matter searches.

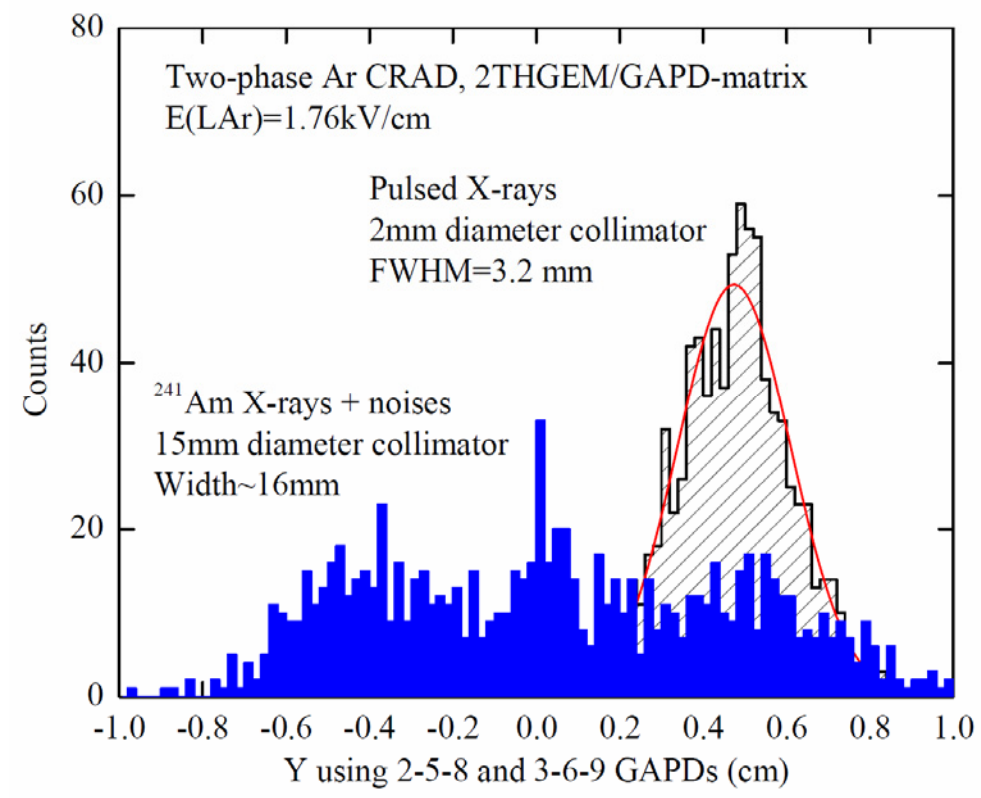

Fig. 6. Y coordinate distribution of the X-ray conversion region in the two-phase Ar CRAD with THGEM/GAPD-matrix optical readout, defined by either 15 mm or 2 mm diameter collimator, for $^{241}$Am X-rays (+ noises) and pulsed X-rays respectively.

We are grateful to A. Chegodaev and R. Snopkov for technical support. This work was supported in part by grants of the Government of Russian Federation (11.G34.31.0047) and the Russian Foundation for Basic Research (12-02-91509-CERN_a and 12-02-12133-ofi_m) and by the Israel Science Foundation grant (477/10).